# Integrating Data Mining and Predictive Modeling Techniques for Enhanced Retail Optimization

Sri darshan M

Department of Artificial Intelligence
and Machine Learning
*SRM UNIVERSITY*
Chennai, India.
sridarshan054@gmail.com

Nithinraj N

Department of Computer Science
*SRM UNIVERSITY*
Chennai, India.
nithin22112004@gmail.com

Jaisachin B

Department of Computer Science
*SRM UNIVERSITY*
Chennai, India.
jaisachin090305@gmail.com

*Abstract*— **Predictive modeling and time-pattern analysis are increasingly critical in this swiftly shifting retail environment to improve operational efficiency and informed decision-making. This paper reports a comprehensive application of state-of-the-art machine learning to the retailing domain with a specific focus on association rule mining, sequential pattern mining, and time-series forecasting. Association rules: Relationship Mining This provides the key product relationships and customer buying patterns that form the basis of individually tailored marketing campaigns. Sequential pattern mining: Using the PrefixSpan algorithm, it identifies frequent sequences of purchasing products-extremely powerful insights into consumer behavior and also better management of the inventories. What is applied for sales trend forecasting models Prophet applies on historical transaction data over seasonality, holidays, and long-term growth. The forecast results allow predicting demand variations, thus helping in proper inventory alignment and avoiding overstocking or understocking of inventory. Our results are checked through the help of metrics like MAE (Mean Absolute Error) and RMSE (Root Mean Squared Error) to ensure our predictions are strong and accurate. We will combine the aspects of all of these techniques to prove how predictive modeling and temporal pattern analysis can help optimize control over inventory, enhance marketing effectiveness, and position retail businesses as they rise to ever greater heights. This entire methodology demonstrates the flexibility with which data-driven strategies can be leveraged to revitalize traditional retailing practices.**

## I. Introduction

In this 21st century, development of technology has played a crucial role in almost every sector, mainly in business. The application of technology in business helps different companies to adopt different strategies on a large scale. The companies are now able to analyze every product and could run their business well. Better market analysis of a company And adopting effective marketing and sales strategies could scale up their sales, thus taking advantage over the rival companies. Data mining helps us to crack the consumer buying patterns in the marketplace. One of the popular data mining techniques That is mainly associated with consumers. Buying patterns. is called market basket analysis.[1-3]. Retailers are now adapted to rapid changes predicting and influencing consumer behaviour. The rise of big data technologies and advancements in the field of artificial intelligence and mission learning has created a path for retailers to maximise the full potential of market basket analysis. However, with the rise in technology, retailers as to face certain challenges like data, privacy, data sources, and the need for skilled personal adapt at interpreting complex, analytical outputs.[4-6]

## II. Market basket analysis

It is one of the shopping cart analysis process which is used in implementing effective marking strategies to meet the products that is to be purchased by the consumers. MBA used in understanding the consumer habits that is used to effectively allocate the stocks based on the sales of a particular product. The study of consumer behaviour is really crucial for businesses to evaluate their marketing strategies and optimise product placements. Over the years, Various powerful analytics has been developed, but one such efficient and powerful method is apirori algorithm with market basket analysis(MBA)[7].

### A. Apirori Algorithm

Apriori algorithm is one of the key components of data mining, extracting methods. The algorithm is an iterative approach that used to replenish the required items based on supply and demand. The Apirori algorithm is a widely used data mining and machine learning techniques Which deals with items in large data sets, i.e. transactional databases.it is an highly efficient algorithm in identifying group of items that occur frequently, so it would be easy for the retailers to replenish the items according to the supply and demand. By using the algorithm, researchers and retailers can extract valuable customer shopping habits, and identify frequent patterns.[8]

### B. Association rule

The association rule mining (ARM) is a mechanism used for calculating Support and confidence of an item relationship. Every single transaction consist of different items, so this method will support the recommendation system based on the supply, demand of the products and finding different patterns in transactions. A study conducted by Xie, The algorithm was used to identify the significant patterns and association rules. Findings help to understand consumer behaviour in a better way, leading to improved Store's efficiency in managing the supply and demand, thus launching targeted marketing campaigns.[9]

### C. FP Growth Algorithm

Frequent pattern growth (FP -Growth) is an algorithm used to determine the frequent item in a dataset. FP-Growth Is developed using Apriori algorithm Frequent pattern growth is one of the algorithm that is a development of A priori algorithm that can be used to determine the most frequent item in the data set. A Study conducted by Chen & Zhang elevated the performance of the FP growth algorithm in market basket analysis, which were really focused on their efficiency, scalability, and is responsible for identifying consumer buying patterns. It also helps the retailers to identify the strength and weakness of each product and its ability to attract the customers. [10-11].

### D. Predictive Modeling

Predictive modeling and pattern analysis become quite important in predicting future trends and supporting informed decisions based on historical data. This section focuses on the methodologies of building predictive models and pattern analyses to enhance retail operations about time-series forecasting and association rule mining. [12]
Predictive modeling uses the historic data in predicting future trends, while pattern analysis aims to find recurrent patterns in the data. These are applied in retail for the prediction of sales, customer behaviour, and requirements for inventory buildup towards the formulation of effective business strategies and making better decisions.[13] Pattern Analysis Pattern analysis techniques are done through association rule mining and sequential pattern mining.[14] The application of the following techniques is as shown:

#### D1. Association Rule Mining:

Association rules are generated using the Apriori algorithm, a method for finding frequent item sets and relations between items. It generates support, confidence, and lift for any given rule and thus characterizes the strength of association.[15]

#### D2. Sequential Pattern Mining:

Apply algorithms like PrefixSpan for sequential pattern identification. This will output the pattern in transaction sequences, indicating what is commonly bought together and when.[10]

#### D3. Predictive Modeling Results:

These models use historic data to project the occurrence of future sales. For example, using an ARIMA model, one can anticipate better sales during holiday seasons. On the other hand, LSTM networks are able to capture seasonal and long-term trends. The training of the models and comparison with actual sales data produces the accuracy of the results.[16]

#### D4. Pattern Analysis Results

Association rule mining brings out the relationships between products, like item pairs commonly purchased together. Sequential pattern mining identifies how often a certain purchase is followed by another purchase, thus letting retailers understand customer behaviour in buying and act accordingly on product placement and promotion.[17]

### E. Temporal Pattern Analysis

Temporal pattern analysis is essential for understanding trends and seasonal behaviors in time-series data. This section explores the methodologies for analyzing temporal patterns in retail transactions, highlighting techniques for hourly and daily distribution analysis, and discussing their implications for inventory management and sales forecasting. Temporal pattern analysis focuses on identifying and understanding patterns within data that vary over time. In the context of retail data, this involves examining how transaction volumes fluctuate across different times of the day and days of the week. Such analyses can provide valuable insights for optimizing inventory management, staff scheduling, and promotional strategies. ss[18-20]

#### E2. Methodology

Temporal pattern analysis is conducted through the following steps:

For the analysis, a synthetic retail dataset is utilized, containing transaction records with timestamps. The dataset includes details such as user ID, transaction ID, item purchased, and timestamp. Data preprocessing involves converting timestamps into datetime objects and extracting relevant features such as the day of the week and hour of the day.

#### E.2.1. Hourly Distribution Analysis:

Transactions are aggregated by hour to determine the distribution of transaction volumes throughout the day. This analysis helps identify peak and off-peak hours, which can inform decisions on staff allocation and operational hours.

#### E.2.2. Daily Distribution Analysis:

Transactions are grouped by day of the week to examine daily patterns. This analysis reveals trends such as increased sales on weekends or weekdays, which can guide promotional strategies and inventory adjustments

### III. IMPLEMENTATION.

The implementation section of this study involves several key steps to follow: synthetic data generation, temporal pattern analysis, association rule mining, sequential pattern mining, and predictive modeling. Each stage is vital for understanding customer behaviors and is used for predicting future trends in retail transactions by the customers.
In this study, we took multi-faceted approach for analyzing and for the prediction of the retail transaction patterns by using the data. This implementation process starts with the generation of a comprehensive synthetic dataset which is designed to emulate real-world retail transactions.

This synthetic data includes essential sections required for this study such as user IDs, transaction IDs, items purchased, and transaction timestamps, covering a period from January 1, to December 31, which is to simulate a realistic retail environment, we generated a synthetic dataset consisting of transactions over a year. Specifically, the dataset generates features of 50 users, 1000 transactions, and 20 distinct items.

This synthetically generated data involves initializing parameters for the number of users, transactions, and items as well as the timestamps of the transactions followed by the creation of a list of item identifiers. For every transaction done, a random user ID and a random number of items (between 1 and 5) are selected from the generated synthetic data list, and the timestamp of the transactions was randomly generated within the defined range of data. This synthetic dataset which is generated allows us to freely explore the various analytical methods without any constraints of real-world data privacy concerns.

We conducted a temporal pattern analysis for understanding of the distribution of transactions over time. By evaluating the distribution of the transactions across different hours of the day and days of the week the temporal patterns were analyzed. The hourly distribution for this analysis involves extracting the hour of the day from transaction timestamps calculates the number of transactions per hour, and visualizes the results with a bar chart. Similarly, daily distribution analysis helped for extracting the day of the week from transaction timestamps and also used for calculating the number of transactions per day.

It identifies the peak shopping hours and days, which is very useful for the retailers to maintain in staffing, inventory management, Promotional activities and to enhance their product selling strategies. This hourly distribution reveals various peaks, indicating high transaction volumes during specific hours by the behavior of customers.

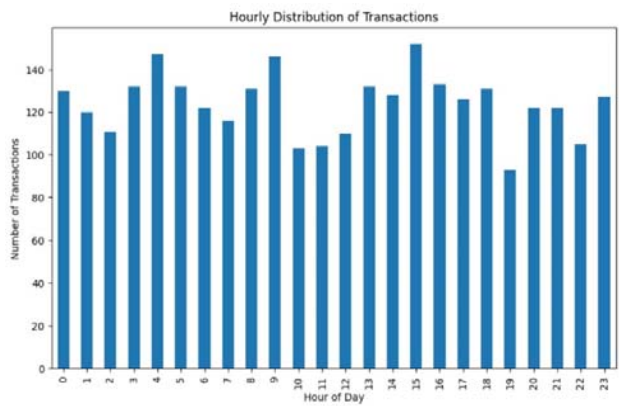

Fig 1.1 Represents the hourly distribution of transactions

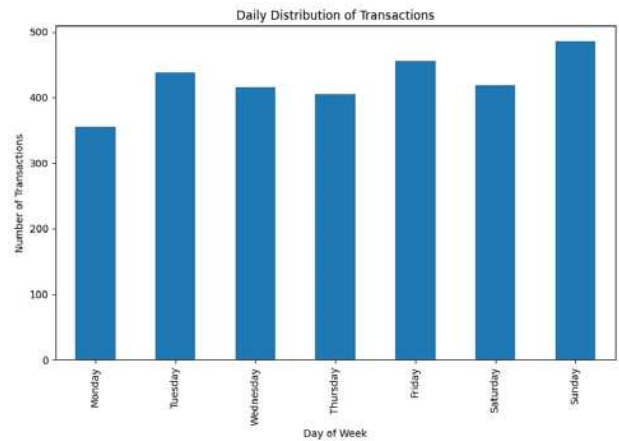

Fig 1.2 Represents the daily distribution of transactions

For the association rule, a mining Apriori algorithm was employed which identifies the frequent itemset and generates association rules that describe how the purchase of one item is associated with the purchase of another. The resulting rules that are generated offer actionable insights for cross-selling and are used for product placement strategies. The analysis of the data based on customer behaviour the model may show that the customers who purchased item2 and item10 will mostly buy the item7 as well (Just an example). The conversion of the transaction data into a one-hot encoded format is done from where the analysis starts, which is tailored for the Apriori algorithm. Frequent item sets are generated with a minimum support threshold of 0.005. From these itemset, association rules were extracted by using the confidence threshold of 0.3, and the resulting rules were analyzed for support, confidence, and lift metrics. Such rules are invaluable for designing targeted marketing campaigns and also enhance the customer's shopping experience.

| | support | confidence | lift | leverage | conviction | zhangs_metric |
|---|---|---|---|---|---|---|
| 0 | 0.006 | 0.315789 | 2.192982 | 0.003264 | 1.251077 | 0.554536 |
| 1 | 0.006 | 0.428571 | 2.747253 | 0.003816 | 1.477000 | 0.645030 |
| 2 | 0.007 | 0.350000 | 2.430556 | 0.004120 | 1.316923 | 0.600583 |
| 3 | 0.006 | 0.375000 | 2.604167 | 0.003696 | 1.369600 | 0.626016 |
| 4 | 0.007 | 0.304348 | 1.938521 | 0.003389 | 1.211812 | 0.495540 |

Fig 1.3 Represents the values produced by apriori algorithm

Sequential pattern mining is another step used for this study, we utilized the PrefixSpan algorithm which helps to identify the frequent sequential patterns in the transaction data. These transactions are then grouped by the columns user and transaction ID to form the ordered sequences of purchased items. This Prefix Span algorithm is further applied to the sequences to find the frequent patterns with a minimum support of 10, and the results are analyzed to encounter the common purchase sequences among customers generated by the synthetic dataset. This analysis helps in the prediction of future purchases based on past behaviors of the customers. We extracted the sequential patterns from the data which highlights the frequent shopping sequences, The customers often buy 'item_17' and followed by the product 'item_18' This is just an example of pattern. These resulting pattens according to the customer behavior help the retailers to manage the inventory and product placements and used to develop their marketing strategies.

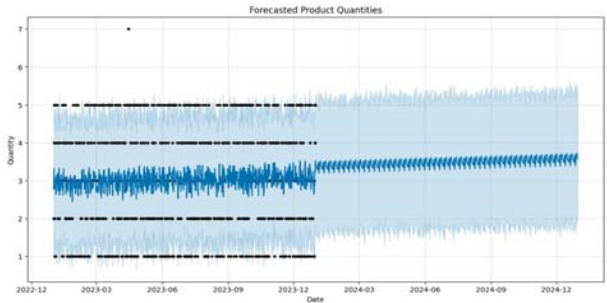

Fig1.4 Represents the change in product quantities

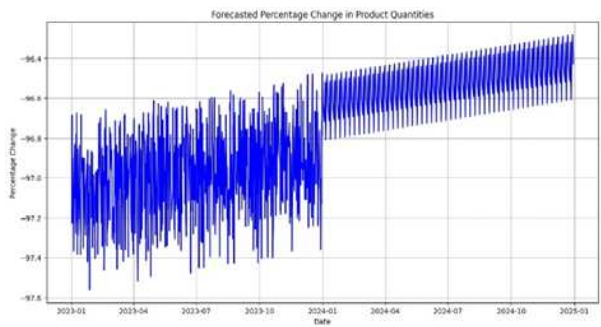

Fig 1.5 Represents the graph from [Fig 1.4].

For predicting the future purchases, we utilized a combined approach which integrates association rules and sequential patterns together. This approach brings the relation between rule antecedents with the items in sequential patterns, thereby creating a set of combined patterns. Based on these combined patterns and the historical transaction data of user's Future purchases are predicted. For instance, if the users purchase history is matching with the antecedents of the association rules, then the items of the rules appear in sequential form, then the consequent items or products are considered as the potential future purchases. By combining both association rules and sequential patterns, we can derive more accurate and precise predictions of future purchases of the customer. This integrated and combined method allows us to understand the behavior of the customers, which helps retailers in making data-driven decisions for the managing of inventories and helping in applying their marketing strategies.

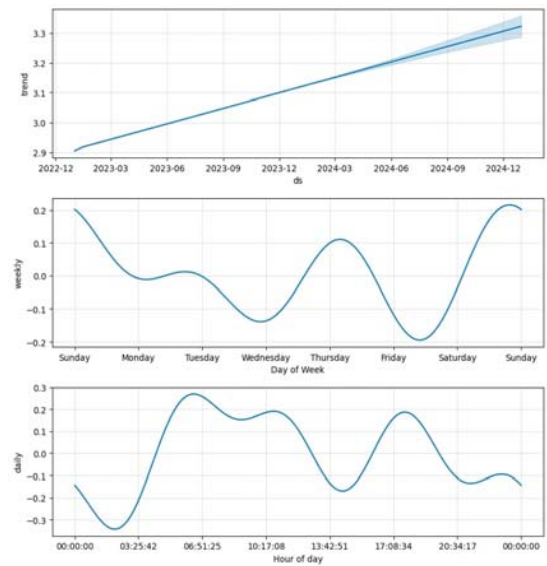

Fig 1.6 Represents the percentage change in quantities.

For this prediction, we used the Prophet model, which is a strong time series forecasting tool used for effective forecasted future quantities of the product. The Prophet model makes predictions on future trends based on historical transaction data, factoring in various components like seasonality, growth, and holiday effects. Focusing on such patterns, the model generates feasible forecasts for the business year and provides an insight into how demand for products would have spiraled with time.

This level of forecast gives a detailed outlook on the percentage change in product volumes over different time horizons for a majority of better-informed decision-making. Through this, business entities can adequately manage the levels of their inventories, hence stock quantities in close alignment with the predicted demands. This will avoid those pitfalls of overstocking, which may lead to waste and increased costs, and understocking, leading to missed opportunities and an unhappy customer base. Furthermore, the forecast insights that are provided by the Prophet model allow businesses to implement inventory methods through strategic planning, which maximizes the quantity of inventory and reduces order lead time, warehouse space use, and efficient sales period inventory. All this CI, data-driven approach ensures that SCM becomes more flexible and responsive, ensuring operational efficiencies with minimal disturbance and boosting customer satisfaction. Better expected demand, justifying a balance between the stock levels, will ensure that the products are in stock when needed without further costs.

Fig 1.7 Represents the overall trends of day and week

The performance of the model is extracted by the usage of metrics such as Mean Absolute Error (MAE), Mean Squared Error (MSE), and Root Mean Squared Error (RMSE). These metrics provide the model 39;s accuracy for which the model is used for predicting future transaction Quantities and provide valuable insights into the effectiveness of the forecasting model.

```
Model Performance Metrics:
Mean Absolute Error (MAE): 1.17
Mean Squared Error (MSE): 1.87
Root Mean Squared Error (RMSE): 1.37
```

## IV. CONCLUSION

This paper illustrates how advanced data mining and predictive modeling can be combined to resolve some of the more complex challenges in retail management. Association rule mining using the Apriori algorithm has helped extract some of the critical associations between products that will drive effective marketing strategies and inventory decisions. Adoption of sequential pattern mining through PrefixSpan resulted in meaningful insights into consumer purchasing behavior that enables the development of very efficient personalized recommendation systems. Moreover, the time-series forecasting capability of the Prophet model has been very instrumental in establishing future trends of sales and maintaining inventory levels with a higher degree of accuracy. The embedding of such methodologies provides a sturdy framework for the optimization of retail operations, most of which tends to prove how data-driven approaches could effectively enhance operational efficiency, customer satisfaction, and overall business performance. Further work could improve these models and apply them in other fields to prove their general applicability and effectiveness.